\documentclass[%
 preprint,
 amsmath,
 amssymb,
 prd,
 longbibliography,
notitlepage
]{revtex4-1}

\usepackage{hyperref}
\usepackage{xspace}
\usepackage[babel]{microtype}
\usepackage{natbib}
\usepackage{etoolbox}

\makeatletter
\g@addto@macro\bfseries{\boldmath}
\makeatother

\makeatletter
\patchcmd{\frontmatter@abstract@produce}
  {\vskip200\p@\@plus1fil
   \penalty-200\relax
   \vskip-200\p@\@plus-1fil}
  {}
  {}
  {}
\makeatother

\usepackage[english]{babel}
\usepackage[T1]{fontenc}
\usepackage[utf8]{inputenc}
\usepackage[scaled=1.04]{biolinum}

\usepackage{fourier}
\usepackage[scaled=0.83]{beramono}

\newcommand{\pvalue}{\emph{p-}value\xspace}

\hypersetup{
    pdfauthor={Andrew Fowlie},
    colorlinks=true,
    allcolors=[rgb]{0.0, 0.2, 0.4},
}

\defcitealias{junk}{``Reproducibility and Replication of Experimental Particle \\ Physics Results''}
\defcitealias{JUNK2}{``Response to Andrew Fowlie’s Comments''}

\begin{document}

\title{Comment on \citetalias{junk}}
\author{\vspace{-0.1cm}Andrew Fowlie\vspace{0.1cm}}
\affiliation{Department of Physics and Institute of Theoretical Physics, Nanjing Normal University, Nanjing, Jiangsu 210023, China}
% hack thanks rather than use email command
\thanks{\textsf{\href{mailto:Andrew.J.Fowlie@NJNU.edu.cn}{Andrew.J.Fowlie@NJNU.edu.cn}}}
%\date{}

\maketitle
\vspace{-0.3cm}
I would like to thank \citet{junk} for beginning a discussion about replication in high-energy physics (HEP). Junk and Lyons ultimately argue that HEP learned its lessons the hard way through past failures and that other fields could learn from our procedures. They emphasize that experimental collaborations would risk their legacies were they to make a type-1 error in a search for new physics and outline the vigilance taken to avoid one, such as data blinding and a strict $5\sigma$ threshold.

The discussion, however, ignores an elephant in the room: there are regularly anomalies in searches for new physics that result in substantial scientific activity but don't replicate with more data. For example, in 2015 ATLAS and CMS showed evidence for a new particle with a mass of about $750\,\text{GeV}$ that decayed into two photons~\citep{digamma_talk}. Whilst the statistical significance was never greater than $5\sigma$~\citep{Khachatryan:2016hje,Aaboud:2016tru}, the results motivated about 500 publications about the new particle, and countless special seminars and talks~\citep{PhysRevLett.116.150001}. The effect did not replicate when the experimental teams analyzed a larger dataset about six months later~\citep{Aaboud:2017yyg,Khachatryan:2016yec}. Although this was a particularly egregious example, experimental anomalies that garner considerable interest before vanishing are annual events~\citep{aps}. 

We are motivated to attempt to control the type-1 error rate because type-1 errors damage our credibility and lead to us squandering our time and resources on spurious effects. Whilst these non-replications aren't strictly type-1 errors as the statistical significance didn't reach the $5\sigma$ threshold and no discoveries were announced, we incur similar damaging consequences, so they cannot be ignored. I shall refer to these errors --- substantial scientific activity including publicly doubting the null and speculating about new effects when the null was in fact true --- as type-1${}^\prime$ errors. Whilst type-1 errors appear to be under control in HEP, type-1${}^\prime$ errors are rampant. 
In the following sections, I discuss these errors in the context statistical practices at the LHC.

\section{Evidence and error rates}
Searches for new physics at the LHC are performed by comparing a \pvalue, $p$, against a pre-specified threshold, $\alpha$.
\begin{samepage}%
There are two common interpretations of this procedure~\citep{doi:10.1198/0003130031856}:\nopagebreak
\begin{enumerate}
    \item\textbf{Error theoretic}~(\citet{10.2307/91247}): By rejecting the null if $p < \alpha$, we ensure a long-run type-1 error rate of $\alpha$. The threshold $\alpha$ specified the desired type-1 error rate and the \pvalue was a means to achieving it.
    \item\textbf{Evidential}~(\citet{fisher}): The \pvalue is a measure of evidence of against the null hypothesis. The threshold $\alpha$ specified a desired level of evidence.
\end{enumerate}
\end{samepage}
Even among adherents of \pvalue{}s, the latter interpretation is considered unwarranted~\citep{doi:10.1177/1745691620958012}, and it is almost never accompanied by a theoretical framework or justification, or a discussion of the desired and actual properties of $p$ as a measure of evidence. 

Unfortunately, Junk and Lyons repeatedly implicitly switch from one to the other. Indeed, the authors interpret $p$ as a measure of evidence and $\alpha$ as a threshold in evidence, e.g., justifying $5\sigma$ by ``extraordinary claims require extraordinary evidence'' and stating that ``[$3\sigma$] or greater constitutes `evidence'.'' We know, however, that interpreted as a measure of evidence, $p$ is incoherent~\citep{doi:10.1080/00031305.1996.10474380,wagenmakers} and usually overstates the evidence against the null~\citep{doi:10.1198/000313001300339950,doi:10.1080/01621459.1987.10478397,berger1987}. For example, there exists a famous bound~\citep{vovk,doi:10.1198/000313001300339950} implying that under mild assumptions $p = 0.05$ corresponds to at least about $30\%$ posterior probability of the null. This was in fact the primary criticism in \citet{rss}. Consequently, one factor in the prevalence of type-1${}^\prime$ errors may be that
\begin{enumerate}
    \item Physicists interpret \pvalue{}s as evidence (as do Junk and Lyons)
    \item Based on \pvalue{}s, physicists overestimate the evidence for new effects
    \item Substantial scientific activity on what turn out to be spurious effects
\end{enumerate}
Unfortunately, \pvalue{}s simply can't give researchers (including Junk and Lyons) what they want --- a measure of evidence --- leading to wishful and misleading interpretations of $p$ as evidence~\citep{cohen}. This cannot be overcome by better statistical training; it is in inherent deficiency of \pvalue{}s and no amount of education about them will imbue them with a coherent evidential meaning.

\section{Controlling errors}

Controlling error rates depends critically on knowing the data collection and analysis plan --- the intentions of the researchers and what statistical tests would be performed under what circumstances --- and adjusting the \pvalue to reflect that. There are, however, an extraordinary number of tests performed by ATLAS, CMS and LHCb at the LHC and elsewhere. This already makes it challenging to interpret a \pvalue{} at all and undoubtedly contributes to the prevalence of type-1${}^\prime$ errors.

Junk and Lyons rightly celebrate the trend in HEP to publicly release datasets and tools for analyzing them. This, however, raises the specter of data dredging. Massive public datasets~\citep{open} combined with recent developments in machine learning~\citep{Kasieczka:2021xcg} could enable dredging at an unprecedented scale. We must think about what precautions we need to prevent misleading inferences being drawn in the future; e.g., pre-registration of planned analyses as a requisite to accessing otherwise open data. Other more radical proposals, to the problems here and elsewhere, include moving away from an error theoretic approach or any approach based on \pvalue{}s.

\section{Final words on \citetalias{JUNK2}}

I would like to thank \citet{JUNK2} for their detailed response to my comments. In the interests of brevity, I respond to only a few of the points (labeled A -- H in \citet{JUNK2}). First, I acknowledge that \citet{junk} aren't mistakenly equating $p$ with the posterior of the null (C). My previous comment is at fault if it implied otherwise. Similarly, by ``evidence'' I assumed that the authors meant nothing more than observations that should change our opinion~\citep{sep-evidence,Morey2016}. The fact that $p$ doesn't equal the posterior of the null is trivial and not that interesting. The fact that $p$ is typically much less than the posterior and that there exist theorems demonstrating it across broad classes of priors isn't trivial and shouldn't be taken lightly (B). 

Second, I find it an over-simplification to say that experimental particle physics usually only considers an error-theoretic interpretation of \pvalue{}s (A). In reality, as in \citet{junk}, our interpretation of $p$ is an ``anonymous hybrid''~\citep{doi:10.1198/0003130031856} of evidential and error theoretic. 
Why else would we respond at all to anomalies below $5\sigma$ if we didn't consider them evidence for new effects? 
% In their response, \citet{JUNK2} attempt to clarify that by ``[$3\sigma$] or greater constitutes `evidence''' they mean that $5\sigma$ is a ``stronger ‘discovery’ claim'' than $3\sigma$. Is a ``stronger ‘discovery’ claim'' something other than one that is supported by greater evidence?

Lastly, there are two elements to consider regarding our responses to anomalies. First, what is the appropriate response from our community to a particular strength of evidence for a new effect? and second, are we misled by $p$ about that strength of evidence, and would we respond differently if we recognized that $p$ typically overstated the evidence or if we were able to easily incorporate the prior plausibility of the new effect?

\bibliography{references}
\end{document}